# Sub-50 nm focusing of 405 nm laser by hemispherical silicon nanolens

## Zhong Wang and Weihua Zhang[*]

*College of Engineering and Applied Sciences, Jiangsu Key Laboratory of Artificial Functional Materials, MOE Key Laboratory of Intelligent Optical Sensing and Manipulation, State Key Laboratory of Analytical Chemistry for Life Sciences, Nanjing University, Nanjing 210093, China.*

*Corresponding author: WH Zhang, zwh@nju.edu.cn

**In this work, we study the light focusing behaviors of sub-micron Si hemispherical nanolens in theory. Results show that the width and depth of the focus spot light at 405 nm can reach 42 nm (approximately $\lambda/10$) and 20 nm ($\lambda/20$), respectively. Theoretical analysis indicates that this nano-focusing phenomenon comes from two reasons, the high refractive index of Si and the sub-micro size of the lens which considerably decrease the influence of material losses. The focusing capability of Si nanolens is comparable with current EUV technique but with a low cost, providing an alternative approach towards super-resolution photolithography and optical microscopy.**

http://dx.doi.org/10.1364/PRJ.xxx

## 1. INTRODUCTION

Photolithography is the backbone of today's information technique industry, and its resolution decides the density, size and performance of transistors. However, due to the wave nature of photons, the resolution (half pitch) of photolithography is fundamentally limited by the diffraction limit, $\Delta x = \lambda/2\text{NA}$ [1]. To push the resolution, in the last few decades, enormous amount of resources has been poured into the field of advanced light sources to shorten the wavelength. Today, commercial successes have been achieved with EUV lithography [2-6], but the cost of EUV based FAB is still prohibitively high. Meanwhile, to circumvent the request for EUV source, number of alternative photolithography techniques were developed including new laser sources [7-9], near-field optical methods [10-13] and plasmonic/metamaterials based lithography [14-18]. But, none of them can achieve a small pitch size and large depth of focus comparable to today's EUV techniques simultaneously.

In addition to the above techniques, Abbe's diffraction limit, $\Delta x = \lambda/2n\sin\theta$, indicates that the resolution can also be improved by immersing the system in a high refractive index material. Indeed, techniques have also been reported based on this idea, including solid immersion lens [19-23], photonic nanojet [24-26], and oil/water immersion technique [27, 28], but their resolutions are ultimately limited by the refractive index of the immersion media. In solid state physics, it is known that high refractive indices are commonly linked with a narrow bandgap because of the Kramer-Kronig relation [29]. As a result, materials with high refractive indices are always lossy in the blue and ultraviolet spectral range. For example, the bandgap of high refractive index materials like Si, Ge and GaAs are all in the infrared regime [30], and they are all opaque in the UV-visible regime. However, interestingly, as an indirect bandgap semiconductor, the absorption of Si is much weaker than direct band semiconductors, and the penetration length can reach hundreds of nanometers, larger than the effective wavelength of light in Si. In other words, it is possible to construct a small/thin Si lens which enjoys the high refractive index without suffering severe material losses. Intrigued by this idea, we study the light focusing performance of Si nanolenses in theory, and investigate the fundamental properties including the full width of half-maximum (FWHM), focus depth and size of the sidelobes.

This paper is organized as follows. First, the light focusing behaviors of sub-micron hemispherical Si nanolenses at 405 nm are studied. Then, the sidelobe depression technique, size-dependent behaviors and polarization effects of Si nanolenses are discussed. Finally, we search the smallest focus spot of Si lens in the whole near-UV spectral range and demonstrate a 35 nm nanofocus spot, which is comparable with the focusing capability of current EUV technique.

## 2. Nanofocusing by silicon nanolens

Figure 1a illustrated the system investigated in this work. A hemispherical Si nano-lens with a radius of *R* is placed in the focus spot of a tightly focused beam produced by an objective lens (N.A. = 0.95). The behaviors of the silicon nanolens (SNL) were calculated using a commercial finite-different time-domain (FDTD) solver (Lumerical FDTD). In details, we first calculated the field distribution at the focused spot of the objective lens without SNL using the Kirchhoff angular spectrum diffraction method [31]. Then, we used the focused fields as the source in FDTD to illuminate the SNL with the help of the total/scattering field boundary technique, and calculated the field distribution in the optical near-filed of the SNL. Perfect matched layer boundary conditions were used to eliminate unwanted reflections at the boundaries; the mesh size was 2 nm, and the refractive index of Si by Palik was used [32]. The wavelength was set to 405 nm (H-line of mercury lamp) in this work, because it is one of the most widely used wavelengths in conventional photolithography. The refractive index of Si is 5.42 at 405 nm, and we can therefore expect a 40 nm focus spot with Abbe's diffraction limit.

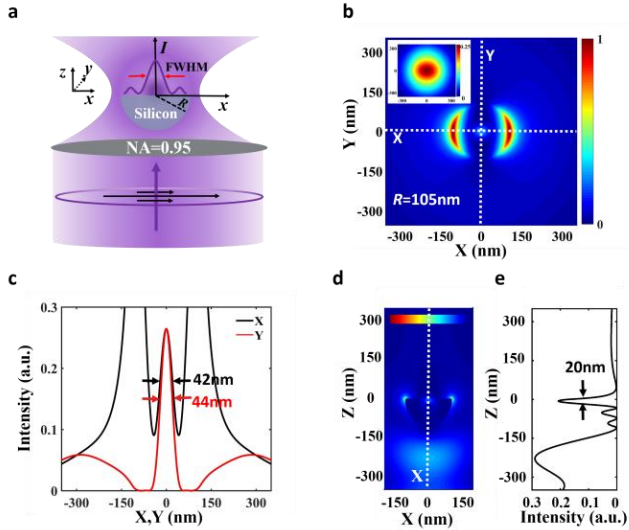

**Fig. 1. a** Schematic drawing of the SNL-based nanofocusing system. **b** Intensity distribution of the nanofocus spot by a SNL ($R$ = 105 nm). The inset shows intensity distribution without SNL. **c** Intensity profiles of the nanofocus spot along the x and y axes labeled in **b**. The FWHM is 42 nm and 44 nm, respectively. **d** Intensity distribution in the x-z plane. **e** Intensity profile along the z axis.

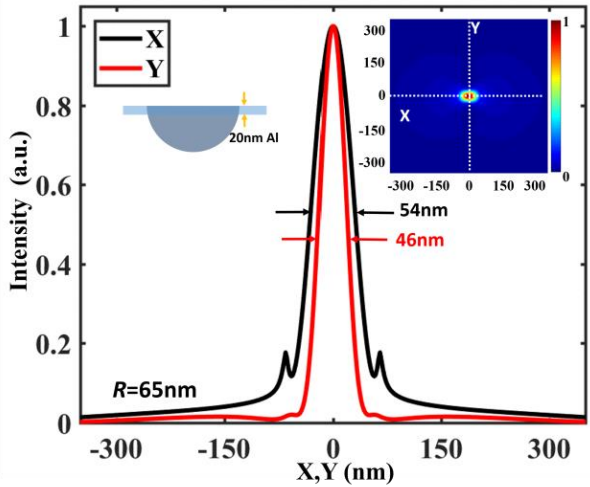

**Fig. 2.** Intensity profile of the nanofocus spot of a SNL with Al mask. The inset shows the intensity distribution in the x-y plane.

Figure 1b shows a typical field distribution of the nano-focused spot by a SNL ($R$ =105 nm) under the illumination of a tightly focused beam with linear polarization at 405 nm. The fields are squeezed by the SNL into a nanometer spot at the center of the focus plane. The FWHM of the nanofocus spot along the x and y axes is 42 nm and 44 nm (Fig. 1c), respectively. These values are approximately 1/10 of the wavelength. The high lateral confinement of light is accompanied with an extremely small depth of focus (DOF). As shown in Fig. 1d and 1e, the DOF is approximately 20 nm, 1/20 of the wavelength. In addition, due to the nano-concentration, the intensity of the nanofocus spot is considerably larger (5.8-fold) than the case without SNL despite the material losses of Si.

One of the major issues of the SNL technique is the large sidelobes which can be several times stronger than the center focus spot (Fig. 1b). To solve the issue, an Al mask (20 nm thick) is added in the system. Numerical results show that the sidelobes can be eliminated without losing the focusing capability of SNL, as shown in Fig. 2. We calculated the performance of the SNL with different sizes, and the smallest FWHM is 54 nm at R = 65 nm. It is slightly larger than the case without the Al film, and this can be understood by considering the decrease of the numerical aperture of the SNL by the additional Al film.

## 3. Nanofocusing of cylindrically polarized light

In addition to the sidelobes, the SNL also has a few other issues, mainly asymmetric focus spot and small DOF. Particularly, the small DOF may lead to severe troubles in photolithography. It means that only a very thin layer of the resist can be exposed, and this will make the following pattern transfer step extremely difficult.

One way to address above issues is to use radially polarized light. It has been reported that the DOF of conventional lens can effective improved by introducing a radially polarized illumination [31, 33-35]. We calculated the focus performance of SNLs with different sizes, and the best results occurs when $R$ = 150 nm. As shown in Fig. 3a and 3b, a perfectly axial symmetric nanofocus spot was obtained, and the FWHM of the focal spot is 60 nm. The DOF is increased to 75 nm, almost 4 times larger than the case of linear polarization. Interestingly, the sidelobes is also considerably smaller (2.2 times smaller than the intensity of the focus spot). This can be contributed to the large lens size here, which blocks the light from the surrounding area.

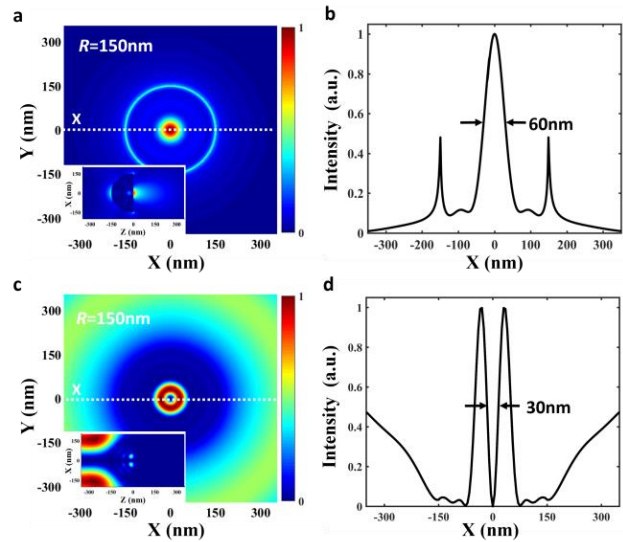

**Fig. 3.** (a) Intensity distribution of the nanofocus spot by a SNL ($R$ = 150 nm) excited by radially polarized light. The inset shows the distribution in the x-z plane. (b) Intensity profiled along the x-axis. (c) Intensity distribution in the case of azimuthally polarized light. The inset shows the x-z plane. (d) Intensity profiled along the x axis.

As a comparison, we also simulated the case of azimuthally polarized light [31, 35-37], another widely used cylindrically polarized beam. The results are shown in Figure 3c and 3d. Similar to the case of conventional objective, a donut-shape spot is observed and the intensity at the symmetric axis is zero. The FWHM of central dark spot is 30 nm, only λ/13. This small size makes this nano-donut beam potentially useful in stimulated depletion emission microscopy and related fabrication techniques [38, 39], in which a donut-shape beam is used for depleting the active molecules in the surrounding area in order to increase the spatial resolution.

## 4. Wavelength dependent behaviors

The nano-focusing capability of SNL comes from the high refractive index of Si and the sub-micrometer size which helps SNLs circumvent the material loss issue. On the other hand, since the size of the SNL is comparable with the wavelength, the conventional ray optics is not applicable here anymore. Instead, we need to consider the nano-focusing behavior as part of the scattering process of the SNL, and this make it fundamentally different from the conventional lens.

As a scattering process, one natural consequence is that the performance of SNL is sensitive to its size, as well as wavelength. And these behaviors can indeed be observed.

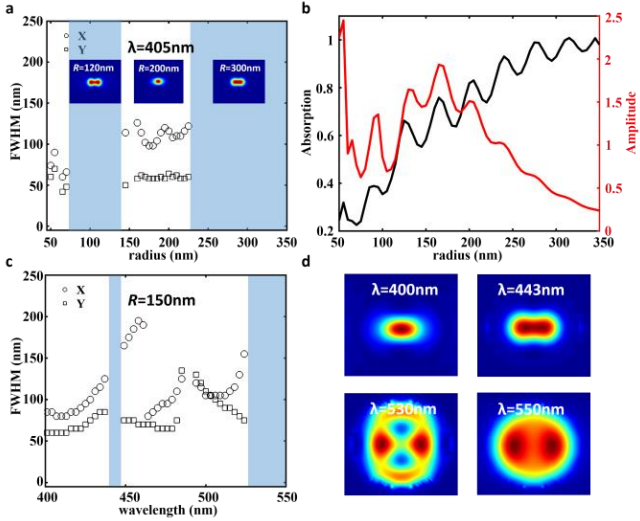

**Fig. 4.** (a) FWHMs of nanofocus spot along the x axis and y axis by SNLs with different radius. Linear polarized light is used. The inset shows some of the typical intensity distribution of SNLs. (b) Absorption and field intensity as a function of the radius of SNL (with Al plate). (c) FWHMs of nanofocus spot by SNL with a fixed radius (150 nm) as a function of wavelength. (d) Typical Intensity distributions of the nanofocus spots by a SNL. Al mask is used in all cases.

We calculated how size influences the optical behaviors of a SNL with an Al mask at 405 nm, and size-dependent behaviors were observed. The absorption of the SNL oscillates with the radius similar to the case of Mie scattering. In addition, the focus behavior of SNL also varies. Nanofocus only occurs in some parts of the size spectrum ($R$ = 50 nm – 70 nm, 140 nm – 230 nm), and the FWHM of the focus spot also varies with $R$. For instance, FWHM = 60 nm when $R$ = 65 nm, and FWHM = 120 nm when $R$ = 200 nm. In the non-nanofocus regime, sophisticated field distributions are observed, e.g. twin-nanospot at $R$ = 120 nm and nanobar at $R$ = 300 nm. When $R$ is larger than 400 nm, the pattern disappears due to the light absorption of Si.

We also investigate the wavelength-dependent behaviors of SNL. Figure 4c and 4d shows the results for the SNL with $R$ = 150 nm. The field distribution at the focus plane is highly wavelength-dependent. Single nanofocus spot can only be observed in part of the spectral range, and sophisticated patterns are found in the rest part, e.g nanobar at 443 nm, ring at 530 nm and twin-spot at 550 nm. Since the supper-resolution focus spot at the focus plane of SNL can be treated as the superposition of the resonant modes of SN, it is not surprising to see this wavelength-dependent behaviros.

## 5. MATERIAL LOSSES

Another important difference between the SNL and conventional lenses is the high material loss of SNL. Conventionally, only high quality transparent materials are used for making lenses in order to have a high transmission efficiency, and how the material loss influence focusing performance of a lens, particularly a nanolens, has not been studied before.

To investigate the effects of material losses, here we change the imaginary part of the refractive index from 0 to 1.5 with the real part, n', fixed at 5.42 (the value of Si at 405 nm), and simulate the change of optical properties of the lens. As shown in Figure 5a, we find that the size of the focus spot stays constant, while the intensity of the focus spot gradually drops with the increasement of n" (material losses). The focus spot starts to vanish when the imaginary part (n") becomes large (in this case, n">1.2). In other words, when the material loss is low, it does not change the field distribution and only induces relative intensity changes. For Si, n" = 0.33 at 405 nm, belongs to the low loss regime, and we can therefore observe the strong nano-focusing effect by the SNL.

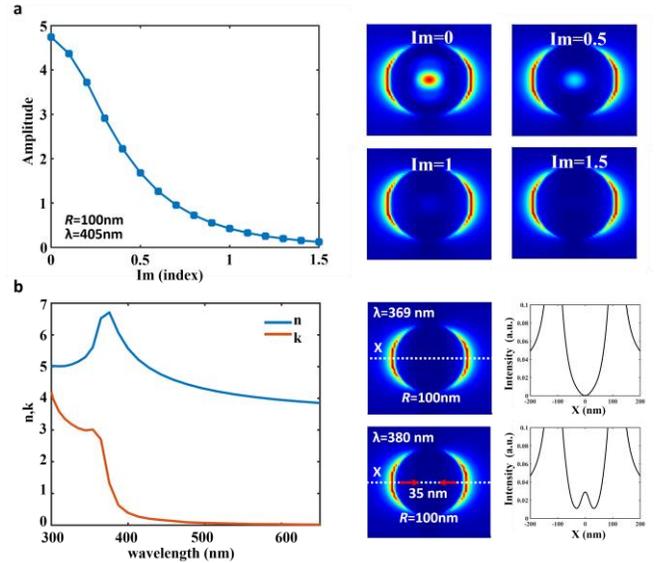

**Fig. 5.** (a) Intensity of the nanofocus as a function of the imaginary part of the refractive index of the SNL. The inset shows field distributions of a SNL with different material losses. (b) Refractive index of silicon at different wavelengths. Intensity distribution of the nanofocus spot of a SNL (R = 100 nm) at 369 nm and 380 nm.

## 6. Nanofocusing in near-UV

In the previous discussion, 405 nm (H-line) is used because it is one of the most commonly used wavelengths for photolithography. However, in theory, shorter wavelength can give a better spatial resolution. Apparently, 405 nm is not the optimal wavelength for nanofocusing of light.

In addition, the real part of refractive index of Si is higher in the near-UV regime and reaches its maximum value (n' = 6.79) at wavelength 369 nm, while the imaginary part also becomes very large (>1) at 369 nm, enough to destroy the focusing capability of the SNL.

In order to find the best focusing result, we simulated the focusing behaviors of SNLs with various diameters at different wavelengths between 369 nm and 405 nm, and found that 380 nm is the optimal wavelength. As shown in figure 5b, the FWHM of a SNL with R=100 nm reaches 35 nm when being excited by a linearly polarized beam at 380 nm. This is even better than the that of current EUV technique. We also calculated the field distribution at 369 nm (Figure 5d), and no focus spot is formed due to the high material losses. This is consistent with the previous discussion.

We summarized the specifications of SNLs, namely wavelength, FWHM of the spot, DOF and sidelobes, as well as those of other major optical nano-focusing techniques today in Table 1. It is evident that the SNL enjoys some key advantages. First, the FWHM of the focus spot, consequently the pitch size in photolithography, can be below 40 nm. This is comparable with EUV technique, and considerably better than the current DUV technique, as well as other focus super resolution enhancement techniques, such as, superoscillatory lens [40-43], solid immersion lens and photonic nanojet. In addition, the SNL uses blue and near-UV light. In this spectral range, there are cheap compact semiconductor lasers available, and they are also convenient to use. On the contrary, the DUV and EUV light sources are extremely expensive, bulky and difficult to use.

Table 1 Reported results for subwavelength focused spots

| Type | FWHM | Depth of Focus | Wavelength | Sidelobes |
|---|---|---|---|---|
| Extreme Ultraviolet (EUV) [2] | 40nm | 248nm | 13.5nm | No |
| Deep Ultraviolet (DUV) [9] | 71nm | 284nm | 193nm | No |
| Super-oscillatory lens (SOL) [42] | 185nm | 7040~60160nm | 640nm | Strong |
| Solid immersion lens (SIL) [19] | 109nm | 300nm | 436nm | No |
| Photonic nanojet (PNJ) [26] | 87nm | 250nm | 632nm | No |
| Silicon nanosphere lens (SNL) | | | | |
| Linearly | 42nm | 20nm | 405nm | Yes |
| Linearly (Al plate) | 54nm | 20nm | 405nm | No |
| Radially | 50nm | 75nm | 405nm | Yes |
| Linearly | 35nm | 15nm | 380 nm | Yes |

Finally, we would like to point out that there are also metal material based super-resolution photolithography techniques, e.g. surface plasmon lithography and metamaterial based photolithography. However, these methods use noble metals, mainly Au, which is toxic to the semiconductor devices. This severely limits their applications in the semiconductor industry.

## 6. SUMMARY

In this work, we studied the light focusing capability of Si nanolens. Numerical results show that at 405 nm (H-line of mercury lamp), the Si nanolens can confine a linearly polarized beam into a 42 nm spot (approx. λ/10) with a DOF of 20 nm (approx. λ/20). In the case of radially polarized light, the DOF can be increased to 75 nm. We also investigated the case of shorter wavelengths, and found that the size of focus spot can even reached 35 nm at 380 nm, which is comparable with the pitch size of current EUV technique. There are two key factors for achieving such nano-focusing capability, namely the high refractive index and the small lens size. In our case, the lens size is comparable or even smaller than the penetration depth of light in Si, and this minimizes the influence of materials losses. We believe that with the nano-focusing capability at visible and near-UV spectral range, the Si nanolens can have many important applications, particularly in the field of super-resolution photolithography and imaging.

**Funding**. This study was supported by grants from the National Key Technologies R&D Program of China (No. 2016YFA0201104).

**Disclosures.** The authors declare no conflicts of interest.